\documentclass[smallextended]{svjour3}       
\smartqed  
\usepackage{graphicx}
\begin{document}

\title{Compatibility of Larmor's Formula with Radiation Reaction for an Accelerated Charge}
\titlerunning{Compatibility of Larmor's Formula with Radiation Reaction}
\author{Ashok K. Singal}
\authorrunning{A. K. Singal} 
\institute{A. K. Singal \at
Astronomy and Astrophysics Division, Physical Research Laboratory,
Navrangpura,\\ Ahmedabad - 380 009, India\\
\email{asingal@prl.res.in}         }
\date{Received: date / Accepted: date}
\maketitle
\begin{abstract}
It is shown that the well-known disparity in classical 
electrodynamics between the power losses calculated from the radiation reaction and that from Larmor's formula,
is succinctly understood when a proper distinction is made between 
quantities expressed in terms of a ``real time'' and those expressed in terms of a retarded time. 
It is explicitly shown that an accelerated charge, taken to be a sphere of vanishingly small radius $r_{\rm o} $, 
experiences at any time a self-force 
proportional to the acceleration it had at a time $r_{\rm o} /c$ earlier, while the 
rate of work done on the charge is obtained by a scalar product of the self-force with the instantaneous (present) 
value of its velocity. Now if the retarded value of acceleration is expressed in terms of the present values of 
acceleration, then we get the rate of work done according to the radiation reaction equation, 
however if we instead express the present value of velocity in terms of its 
time-retarded value, then we get back the familiar Larmor's radiation formula.
From this simple relation between the two we show that they differ because Larmor's formula, in contrast with the radiation reaction, 
is written not in terms of the real-time values of quantities specifying the charge motion but is instead 
expressed in terms of the time-retarded values. Moreover, it is explicitly shown that the difference in the two formulas for 
radiative power loss exactly 
matches the difference in the temporal rate of the change of energy in the self-fields between the retarded and real times. 
From this it becomes obvious that the ad hoc introduction 
of an acceleration-dependent energy term, usually referred to in the prevalent literature as Schott-term, 
in order to make the two formulas comply with each other, is redundant.
\end{abstract}
\section{Introduction}
One of the most curious and perhaps an equally annoying problem in 
classical electrodynamics is that the power emitted from an 
accelerated charge does not appear to conform with the
radiation reaction on the charge. In the standard,
Larmor's radiation formula (generalized to Li\'{e}nard's
formula in the case of a relativistic motion), the radiated
power is directly proportional to the {\em square of acceleration}
of the charged particle. From the energy conservation 
law it is to be surely expected that the 
power emitted in radiation fields equals the power loss undergone 
by the accelerated charge. But from the radiation reaction equation 
the power loss of an accelerated charge is directly proportional
to its {\em rate of change of acceleration} \cite{1,2,3}.
Although the two formulas do yield the same value
of energy when integrated over a time interval chosen such that 
the scalar product of velocity and acceleration vectors of the 
charge is the same at the beginning as at the end of the interval 
(a periodic motion of an oscillating charge or a circular motion like in 
synchrotron case)~\cite{52,48}, the two 
calculations do not match when the charge is still undergoing
a non-uniform, non-periodic motion at either end of the time 
interval~\cite{1}. In any case the functional forms of the two formulas
appear quite different. This enigma has defied a 
satisfactory solution despite the continuous efforts for the last 
100 years or so. It is generally thought that the root-cause of this
problem may lie in the radiation-reaction equation 
whose derivation is considered to be not as rigorous as that of
the formula for power radiated. 
Some interesting proposals for the ``removal'' of the above discrepancy include 
the ad hoc assumption of an acceleration dependent term either 
in a modified form of the Lorentz-force formula~\cite{49}, or in the 
radiation-reaction equation in the form of an ``acceleration-energy'' term
for an accelerated charge \cite{5,6,44,41} (also called as an 
internal-energy term or simply a ``Schott-term'', based on the first 
such suggestion by Schott \cite{7}), or a somewhat related
proposition \cite{8} that even the {\em proper-mass} of an accelerated 
charged particle (e.g., of an electron) varies, or even a 
combination of some of these propositions \cite{9}. 
Alternately it has been suggested \cite{10} that
there may be some fundamental difference in the electrodynamics of a
continuous charge distribution and of a ``point charge'' (with a 
somewhat different stress-energy tensor for the latter).
It is interesting to note that an understanding
of this anomaly has sometimes been sought beyond the border of the
classical electrodynamics ( e.g., in the vacuum-fluctuations of the 
electromagnetic fields in quantum theory \cite{11}). 
Ideally one would expect the classical electrodynamics to be
mathematically consistent {\em within itself}, 
even though it might not be adequate to explain all experimental 
phenomena observed for an elementary charged particle. 
Because of the vastness of literature on this subject, we refer
the reader to review articles or text-books \cite{12,51,20,21,56,57} for further references on 
these and other interesting ideas that have been proposed to remove 
this seemingly inconsistency in classical electrodynamics.

We intend to show here that the difference perceived in the 
two power formulas is merely a reflection of the fact that the 
two are calculated in terms of two different 
time systems. While the radiation-reaction formula is expressed 
in terms of the acceleration and its temporal derivative, being evaluated at the ``real time'' of the 
charged particle, Larmor's radiation formula is 
written in terms of quantities describing motion of the charge, actually at a retarded time. 
The difference between the two time systems is only $\sim r_{\rm o} /c$ for 
a charged particle of radius $r_{\rm o} $, and is as such vanishingly small for a
charge distribution that reduces to a ``point'' in the limit.
But as we will show below, it still gives rise to a finite apparent difference, 
independent of $r_{\rm o} $, for the power calculations in the two formulas,
because of the presence of the $1/r_{\rm o} $ term in the self-field
energy of the charge. By bringing out this simple relation between 
these two seemingly contradictory results, we demonstrate
in this way their mutual compliance, without invoking any additional hypothesis. 
We shall initially confine ourselves to a nonrelativistic case, laying bare 
the basic ideas, and then using the condition of relativistic covariance (see e.g.,~\cite{2}), 
the formulation would be generalized to a relativistic case.
\section{Larmor's Formula for Power Radiated by a Charge}
The electromagnetic field (${\bf E},{\bf B}$) of an arbitrarily moving charge $e$ at a time $t$ is given by \cite{1,28},
\begin{equation}
\label{eq:1a}
{\bf E}=\left[\frac{e({\bf n}-{\bf v}/c)}{r^{2}\gamma ^{2}(1-{\bf n}\cdot{\bf v}/c)^{3}} + 
\frac{e{\bf n}\times\{({\bf n}-{\bf v}/c)\times \dot{\bf v}\}}{rc^2\:(1-{\bf n}\cdot {\bf v}/c)^{3}}\right]_{t_{\rm o}}\: ,
\end{equation}
\begin{equation}
\label{eq:1b}
{\bf B}=[{\bf n}]_{t_{\rm o}} \times {\bf E}\:,
\end{equation}
where the quantities in square brackets on the right hand side are to be evaluated 
at a retarded time $t_{\rm o}=t-r/c$. More specifically, ${\bf v}$, $\dot{\bf v}$, and 
$\gamma=1/\surd(1-v^2/c^2)$ represent respectively the velocity, acceleration and the Lorentz
factor of the charge at the retarded time, while ${\bf r}=r{\bf n}$ is the radial vector from the retarded 
position of the charge to the field point where electromagnetic fields are being evaluated.

Let the charge be instantly stationary (i.e., ${\bf v}_{\rm o}=0$, though it may have a finite acceleration $\dot{\bf v}_{\rm o} $), 
at a time $t_{\rm o}$, then we can write the electromagnetic fields as \cite{28},
\begin{equation}
\label{eq:1a1}
{\bf E}=\left[\frac{e\:{\bf n}}{r^{2}} + \frac{e\:{\bf n}\times\{{\bf n}\times \dot{\bf v}_{\rm o} \}}{rc^2}\right]_{t_{\rm o}}
\end{equation}
\begin{equation}
\label{eq:1b1}
{\bf B}=\left[\frac{-\:e\:{\bf n}\times\dot{\bf v}_{\rm o} }{rc^2}\right]_{t_{\rm o}} \:.
\end{equation}
The simple relations ~(\ref{eq:1a1}) and ~(\ref{eq:1b1}) do not give the electromagnetic field for all field points 
at all times. Instead these yield the electromagnetic field for events in space-time causally connected to the 
charge position at $t_{\rm o}$ when it had ${\bf v}_{\rm o}=0$ and an acceleration $\dot{\bf v}_{\rm o}$. 
Thus Eqs.~(\ref{eq:1a1}) and ~(\ref{eq:1b1}) give for any time $t=t_{\rm o}+\tau$ the 
electromagnetic field on a spherical surface $\Sigma$ of radius $r=c \tau$ centered on the charge position at $t=t_{\rm o}$. 
Then from Poynting vector \cite{1}, 
\begin{equation}
\label{eq:8}
{\cal S}= \frac{c}{4\pi}{\bf E}\times {\bf B}\:,
\end{equation}
one gets the electromagnetic power passing through $\Sigma$ at $t=t_{\rm o}+\tau$, 
\begin{equation}
\label{eq:1c}
P= \oint_{\Sigma}{{\rm d}\Sigma}\:({\bf n} \cdot {\cal S})
=\frac{e^2[\dot{\bf v}_{\rm o}^2]_{t_{\rm o}}}{2 c^3}\int_{\rm o}^{\pi} {\rm d}\theta\: \sin^3\theta
=\frac{2e^{2}}{3c^{3}} \left[\dot{\bf v}_{\rm o} ^{2}\right]_{t_{\rm o}} \:.
\end{equation}
The power passing through the spherical surface is independent of its radius $r$,  which could be made  
vanishingly small around the charge position at $t=t_{\rm o}$, and then intuitively, from the causality argument, one  
concludes that this must be the energy loss rate of the charge at $t=t_{\rm o}$. This is Larmor's famous result for an 
accelerated charge that the power loss at any time is proportional to the square of its acceleration at that instant.
\section{The Calculation of Self-Force for an Accelerated Charge}
Poynting's theorem allows us to relate the rate of electromagnetic
energy outflow through the surface boundary of a charge distribution to the
rate of change in the mechanical energy of the enclosed charges due to the 
electromagnetic fields within the volume. In our case here, the electromagnetic 
field within the volume is that of the charge itself, so for the charge to lose energy 
its self-field must cause some net force on it. 
For simplicity we shall consider here for the charge particle, a classical uniform spherical-shell model 
of radius $r_{\rm o} $, which may be made vanishingly small in the limit. We need to consider the force 
on each infinitesimal element of the charged sphere
due to the fields from all its remainder parts, with the 
positions, velocities and accelerations of the latter calculated at the retarded times 
(\`{a} la expression~(\ref{eq:1a})).
Then the net force on the charge is calculated by an integration over the whole sphere. 
For a simplification, the force calculations are usually done in the instantaneous rest-frame
of the charge, say, at $t=t_{\rm o}$.

Actually the mathematical details of the calculations of 
self-force, carried out first by Lorentz~\cite{16} and done later more
meticulously by Schott~\cite{7}, 
are available as series in powers of $r_{\rm o} $, in various forms 
in more modern text-books \cite{1,2,3,20}. In such calculations 
it is generally assumed that the motion of the charged particle is non-relativistic and it
varies slowly so that during the light-travel time
across the particle, any changes in its
velocity, acceleration and other higher time derivatives
are relatively small. This is equivalent to the conditions
that $|{\bf v}|\,\ll\,c, \:|\dot{\bf v}|\tau_{\rm o} \,\ll\,c, \:|\ddot{\bf v}|\tau^2_{\rm o} \,\ll 
\,c,$ etc., where $\tau_{\rm o} = r_{\rm o} /c$. Therefore we keep only 
linear terms ${\bf v}/c$, $\dot{\bf v}\tau_{\rm o}/c$, $\ddot{\bf v}\tau^2_{\rm o}/c$, etc., in the formulation of 
self-force. 

Alternately, one can calculate the electric field outside as well as inside of an 
instantaneously stationary charged spherical shell. For a permanently stationary charged sphere, 
it is a simple radial Coulomb electric field outside the shell,
\begin{equation}
\label{eq:2a}
{\bf E}= \frac{e\:{\bf n}}{r^{2}}\:,
\end{equation} 
while the field inside the shell is zero. The above Coulomb field gives rise to an outward repulsive force 
on each charge element of the shell. However, the net force is zero due to the spherical symmetry (we ignore here the 
question of the stability of the charged sphere against the force of self-repulsion). The magnetic field is zero everywhere.
\begin{equation}
\label{eq:1b2}
{\bf B}=0 \:.
\end{equation}
 
However for an accelerated charge there is an additional acceleration--dependent field component   
which gives rise to a non-zero electric field inside the shell \cite{24},
\begin{equation}
\label{eq:2}
{\bf E}=e \left[-\frac{2\dot{\bf v}}{3r_{\rm o} c^{2}}+\frac{2\ddot{\bf v}}{3c^{3}}+\cdots\right]\:,
\end{equation}
while just outside the surface the field is,
\begin{equation}
\label{eq:2.1}
{\bf E}=e \left[\frac{{\bf n}  }{r_{\rm o} ^{2}}-\frac{2\dot{\bf v}}{3r_{\rm o} c^{2}}+\frac{2\ddot{\bf v}}{3c^{3}}+\cdots\right]\:,
\end{equation}
where ${\bf n}  ={\bf r}_{\rm o} /r_{\rm o} $ is the outward radial unit vector at the surface of the sphere of radius $r_{\rm o} $.
Except for the first (Coulomb field) term, the electric field is continuous across the surface and is constant, to this order,
both in direction and magnitude, at all points on the charged sphere. The magnetic field both inside and on the surface 
of the sphere at this instant (i.e., at $t=t_{\rm o}$) is given by \cite{20},  
\begin{equation}
\label{eq:2b1}
{\bf B}=\frac{-\:e\:{\bf n}  \times\ddot{\bf v}}{3 c^3}+\cdots\:.
\end{equation}

Due to the electric field there is a self-force on the charge,
\begin{equation}
\label{eq:3a}
{\bf f}=-\frac{2e^{2}}{3r_{\rm o} c^{2}}\dot{\bf v}+\frac{2e^{2}}{3c^{3}}\ddot{\bf v}\:.
\end{equation}
Here we have dropped terms of order $r_{\rm o} $ or higher, which will become zero for a vanishingly small $r_{\rm o} $. 
This self-force can be written as,
\begin{equation}
\label{eq:3a1}
{\bf f}= -\frac{4U_{\rm el}}{3c^{2}}\dot{\bf v}+\frac{2e^{2}}{3c^{3}}\ddot{\bf v}
= -m_{\rm el}\:\dot{\bf v}+\frac{2e^{2}}{3c^{3}}\ddot{\bf v}\:,
\end{equation}
where $U_{\rm el}=e^{2}/2r_{\rm o} $ represents the electromagnetic self-energy in Coulomb fields of a
stationary spherical-shell charge and $m_{\rm el}=4U_{\rm el}/3c^{2}$ the inertial mass because of 
the electromagnetic self-energy of the charge \cite{29}.

\section{The Rate of Work Done Against the Self-Force}
To calculate the rate of work being done 
against the self-force of a moving charge in an inertial frame,
one has to take the scalar product of 
the self-force ${\bf f}$ and the instantaneous velocity vector ${\bf v}$ of the charge,
both measured in that frame. For a non-relativistic
motion the expression for force can be used directly from that 
in the instantaneous rest frame (Eq.~(\ref{eq:3a1}) above), and if need be, then 
use the condition of relativistic covariance \cite{2,3} to get the more
general formulas valid for any inertial frame which we shall do in later sections.

Accordingly, for an accelerated charge the rate of work done by the charge against self-fields or the corresponding rate of decrease of the 
mechanical energy $({\cal E_{\rm me}})$ of the charge is given by,
\begin{equation}
\label{eq:6}
\frac{{\rm d}{\cal W}}{{\rm d}t}=-\frac{{\rm d}{\cal E_{\rm me}}}{{\rm d}t}=\frac{2e^{2}}{3r_{\rm o} c^{2}}\dot{\bf v}\cdot{\bf v}
-\frac{2e^{2}}{3c^{3}}\ddot{\bf v}\cdot{\bf v}\:,
\end{equation}
or
\begin{equation}
\label{eq:7}
\frac{{\rm d}{\cal W}}{{\rm d}t}= -\frac{{\rm d}{\cal E_{\rm me}}}{{\rm d}t}=
m_{\rm el}\:\dot{\bf v}\cdot{\bf v}-\frac{2e^{2}}{3c^{3}}\ddot{\bf v}\cdot{\bf v}\:.
\end{equation}
The first term on the right hand side in Eq.~(\ref{eq:7}) represents the rate of change of the self-Coulomb field 
energy of the charge as its velocity changes. This term when combined
with the additional work done during a changing Lorentz contraction 
(not included here in Eq.~(\ref{eq:7})) against the Coulomb self-repulsion 
force of the charged particle,
on integration leads to the correct expression for energy in fields 
of a uniformly moving charged particle~\cite{15}. Thus it is only the 
second term ($\propto -\ddot{\bf v}\cdot{\bf v}$) on the right hand side of Eq.~(\ref{eq:7})
that represents the ``excess'' power going into the electromagnetic fields of a charge with a non-uniform
motion. This conforms with the conclusions arrived at otherwise 
that a uniformly accelerated charge does not radiate \cite{17,18}. 

It has always seemed enigmatic that if Larmor's
formula indeed represents the instantaneous rate of power 
loss for an accelerated charge, why the above term contains 
$-\ddot{\bf v}\cdot{\bf v}$ instead of $\dot{\bf v}^{2}$ (cf. Eq.~(\ref{eq:1c})).
It should be noted that though the term containing $-\ddot{\bf v}\cdot{\bf v}$ in Eq.~(\ref{eq:7})) is independent of radius 
$r_{\rm o} $ of the sphere, yet it was derived for a charged sphere (albeit of a vanishingly small radius), 
while Eq.~(\ref{eq:1c}) is for a ``point'' charge. But that could not be the source of the 
discrepancy as from an approximate analysis of the electromagnetic force on a charged spherical shell of a vanishing small radius, 
due to a comoving equivalent ``point charge'' at the centre of the sphere in its instantaneous rest-frame, 
it has been shown that the force due to the time-retarded fields of the point charge equals that due to the time-retarded 
self-fields of the charged sphere itself \cite{50}. This of course should not come as a surprise since the force on a point charge 
at the centre, due to the net electric field of all elements of the spherical charge (Eq.~(\ref{eq:2})), has to be the same as that of 
the point charge on the sphere since both calculations involve exactly the same time-retardation effects when pair-wise diagonally opposite 
infinitesimal element are considered. 
In fact the above could be shown even better from the more rigorously derived expressions for the 
electric field of a point charge in its instantaneous rest frame \cite{24}. Moreover, recently it has been explicitly shown 
that Eqs.~(\ref{eq:3a}) and (\ref{eq:6}) represent the non-relativistic expressions for 
the radiation reaction and the consequential radiative power loss for an arbitrarily moving ``point charge'' \cite{31a,31b}.
In any case, as we shall show below, Larmor's result (Eq.~(\ref{eq:1c})) derived for a point charge is 
indeed consistent with the work done against the self-forces (Eqs.~(\ref{eq:6}) and (\ref{eq:7})),
derived for a spherical charged-shell of vanishing small radius, as long as one understands that the former is expressed in terms of 
the time-retarded motion of the charge while the latter is written in terms of the real-time values of the charge motion.
\section{Larmor's Formula for Radiated Power vs. Power Loss due to Radiation Reaction}
In the inertial frame, where the charge was at rest at $t={t_{\rm o}}$, the rate of energy flow through surface $\Sigma$ is given by 
Eq.~(\ref{eq:1c}). That is, the radiation passing through the spherical surface at $t={t_{\rm o}}+r/c$ involves 
$\dot{\bf v}_{\rm o} $, the acceleration evaluated at retarded time $t={t_{\rm o}}$. This electromagnetic power crossing the surface $\Sigma$ 
at time $t={t_{\rm o}}+ r/c$ was equated to the energy loss rate of the charge at $t={t_{\rm o}}$ to arrive at Larmor's formula. 
There is something amiss here otherwise how could the charge have undergone any power loss 
at $t={t_{\rm o}}$ since it had no kinetic energy  at that instant that could have been lost? 
One could perhaps justifiably argue that the external agency responsible for maintaining the presumed acceleration of the charge in spite of the 
radiative losses, might be the ultimate source of radiated power, but here it could not have provided the power necessary for radiation 
since work done by this external agency too will be zero at that instant as the charge has a zero velocity. Moreover, equating the 
Poynting flux through a surface at a time ${t_{\rm o}}+\tau$ to the loss rate of the mechanical energy of the charge {\em at another instant}
(viz. $t={t_{\rm o}}$) is not a mathematically correct approach, because Poynting's theorem relates the electromagnetic power crossing a closed  
surface to the energy changes within the enclosed volume {\em at the same instant} only. 
Poynting's theorem for an electromagnetic system states that at any particular instant
the rate of energy flow out of a surface plus the time rate of increase of electromagnetic 
energy within the enclosed volume is equal to negative of the mechanical work done by the field 
on the charges within the volume \cite{1,25}. Therefore the sum total of the 
rate of change of the volume integral 
of electromagnetic field energy $({\cal E_{\rm em}})$ and mechanical energy $({\cal E_{\rm me}})$ of the charge 
within the enclosed volume, {\em evaluated only at} $t={t_{\rm o}}+r/c$, should be equated to the negative of 
the surface integral of the Poynting flow in Eq.~(\ref{eq:1c}). 
\begin{equation}
\label{eq:21a}
\frac{{\rm d}{\cal E_{\rm me}}}{{\rm d}t}+\frac{{\rm d}{\cal E_{\rm em}}}{{\rm d}t}=-\frac{2e^{2}}{3c^{3}} \:\dot{\bf v}_{\rm o} ^{2} \:.
\end{equation}
We calculate the volume integral of the electromagnetic field energy ${\cal E_{\rm em}}$ from Eq.~(\ref{eq:2}) within the charged sphere, 
and find it to be proportional 
to $r_{\rm o} $ (or its higher powers) that can thence be neglected for a vanishing small $r_{\rm o} $, 
implying ${\rm d}{\cal E_{\rm em}}/{\rm d}t=0$. 
Therefore applying Poynting's theorem (Eq.~(\ref{eq:21a})) to the sphere of radius $r_{\rm o} $, we get for the mechanical work done on the 
charged sphere,
\begin{equation}
\label{eq:21b}
\frac{{\rm d}{\cal E_{\rm me}}}{{\rm d}t}=-\frac{2e^{2}}{3c^{3}} [\dot{\bf v}_{\rm o} \cdot \dot{\bf v}_{\rm o} ]\:.
\end{equation}
Now while the power on either side in Eq.~(\ref{eq:21b}) is for $t={t_{\rm o}}+\tau_{\rm o}$, the right hand side is expressed in terms of 
$[\dot{\bf v}_{\rm o}] $, the value of the acceleration at retarded time $t={t_{\rm o}}$. However, we can write the acceleration in terms of its value at time 
$t={t_{\rm o}}+\tau_{\rm o}$ as,
\begin{equation}
\label{eq:22}
[\dot{\bf v}_{\rm o}] =\dot{\bf v}-{\ddot{\bf v}}\tau_{\rm o}+\cdots. 
\end{equation}
Further, the charge at time $t={t_{\rm o}}+\tau_{\rm o}$ is moving with a velocity ${\bf v}$ given by,
\begin{equation}
\label{eq:21}
{\bf v}=[{\dot{\bf v}_{\rm o} }\tau_{\rm o}+\cdots],
\end{equation}
where $\tau_{\rm o}=r_{\rm o} /c$ and we dropped terms of order $\tau_{\rm o}^2$ or higher for a small $r_{\rm o} $. 
Substituting the values of $[{\dot{\bf v}}_{\rm o}]$ from Eqs.~(\ref{eq:22}) and (\ref{eq:21}) 
into Eq.~(\ref{eq:21b}), we can write for the rate of change of the energy of the charge,
\begin{equation}
\label{eq:22b}
\frac{{\rm d}{\cal E_{\rm me}}}{{\rm d}t}=- \frac{2e^{2}}{3c^{3}}(\dot{\bf v}-{\ddot{\bf v}\tau_{\rm o}})\cdot\frac{\bf v}{\tau_{\rm o}}\:,
\end{equation}
or
\begin{equation}
\label{eq:22c}
\frac{{\rm d}{\cal E_{\rm me}}}{{\rm d}t}= -\frac{2e^{2}}{3r_{\rm o} c^{2}}\dot{\bf v}\cdot{\bf v}
+\frac{2e^{2}}{3c^{3}}\ddot{\bf v}\cdot{\bf v} \:. 
\end{equation}
Thus we see that the rate of change of the energy of the charge (Eq.~(\ref{eq:22c})), 
inferred from the Larmor's radiation formula (Eq.~(\ref{eq:1c})),
is equal and opposite to the rate of work done (Eq.~(\ref{eq:6})) by the charge against the self-fields.

With the hindsight one can understand the radiation-reaction equation now more clearly. 
Using Eq.~(\ref{eq:22}), we can write the self-electric field in Eq.~(\ref{eq:2}) as,
\begin{equation}
\label{eq:2d}
{\bf E}= -\frac{2e}{3r_{\rm o} c^{2}}[\dot{\bf v}_{\rm o}] \:,
\end{equation}
and thereby the force on the charge from Eq.~(\ref{eq:3a}) as,
\begin{equation}
\label{eq:3d}
{\bf f}=-\frac{2e^{2}}{3r_{\rm o} c^{2}}[\dot{\bf v}_{\rm o}] = -m_{\rm el}\:[\dot{\bf v}_{\rm o}] \:,
\end{equation}
and accordingly a rate of work done,  
\begin{equation}
\label{eq:7a}
\frac{{\rm d}{\cal W}}{{\rm d}t}= -\frac{{\rm d}{\cal E_{\rm me}}}{{\rm d}t}=\frac{2e^{2}}{3r_{\rm o} c^{2}}[\dot{\bf v}_{\rm o}] \cdot{\bf v}
=m_{\rm el}\:[\dot{\bf v}_{\rm o}] \cdot{\bf v}=\frac{2e^{2}}{3c^3}[\dot{\bf v}_{\rm o}^2]\:,
\end{equation}
where we made use of Eq.~(\ref{eq:21}) to get Larmor's formula (Eq.~(\ref{eq:1c})).

We can also examine the consistency of radiated power to that of nil rate of work being done against 
the self-forces of the charge at time $t_{\rm o}$, because the charge has a zero instantaneous velocity. Actually to do that we have to apply the  
Poynting theorem for that very instant $t_{\rm o}$ and therefore calculate the Poynting flux  also through the spherical surface at 
${t_{\rm o}}$ (which is not the same as calculated in Eq.~(\ref{eq:1c}) as there the Poynting flux calculated was for time 
$t={t_{\rm o}}+r_{\rm o}/c$). 
From electric and magnetic fields on the surface of the charge (Eqs.~(\ref{eq:2.1}) and (\ref{eq:2b1})) at time ${t_{\rm o}}$, we can write 
\begin{equation}
\label{eq:8.1}
{\cal S}= \frac{-e^2}{6\pi}\left[\frac{{\bf n}  \times({\bf n}  \times \ddot{\bf v})}{2r_{\rm o} ^2 c^2}
-\frac{\dot{\bf v}\times({\bf n}  \times \ddot{\bf v})}{3r_{\rm o} c^4}
+\frac{\ddot{\bf v}\times ({\bf n}  \times \ddot{\bf v})}{3 c^5}\right].
\end{equation}
\begin{equation}
\label{eq:8.2}
P= \frac{e^2}{6\pi}\oint_{\Omega} r_{\rm o}^2 \:{\rm d} \Omega\:{\bf n}\: \cdot 
\left[\frac{\dot{\bf v}\times({\bf n}  \times \ddot{\bf v})}{3r_{\rm o} c^4}
-\frac{\ddot{\bf v}\times ({\bf n}  \times \ddot{\bf v})}{3 c^5}\right].
\end{equation}
It can be readily seen that the net Poynting flux through the surface at $t={t_{\rm o}}$ would have terms only of order $r_{\rm o}$ 
or higher, which would vanish for a vanishingly small $r_{\rm o}$. There is no radiative flux term which is independent of the 
radius of the sphere, as would be expected in Larmor's formula (c.f. Eq.~(\ref{eq:1c})). This nil radiative loss rate at $t={t_{\rm o}}$ 
(i.e., when the charge is instantaneous stationary with a nil rate of work being done against its self-forces from Eq.~(\ref{eq:7})), 
demonstrates that the two methods of calculating power are indeed compatible with each other.
\section{Redundancy of the Acceleration--Dependent Energy Term}
As shown above (Eq.~(\ref{eq:3d})), an accelerated charge experiences a self-force proportional to the acceleration it 
had at a time interval $\tau_{\rm o}=r_{\rm o} /c$ earlier (basically due to the finite speed $c$ of the electromagnetic field propagation).
In other words, the self-force on the charge at time $t$ is proportional to the acceleration at a retarded time $t_{\rm o}=t-r_{\rm o} /c$. 
\begin{equation}
\label{eq:3d1}
{\bf f}_t=-\frac{2e^{2}}{3r_{\rm o} c^{2}}[\dot{\bf v}]_{t_{\rm o}} = -m_{\rm el}\:[\dot{\bf v}]_{t_{\rm o}}\:.
\end{equation}
Then the rate of work done by a moving charge against the self-force at time $t$ is 
proportional {\em not} to the ``real time'' value of the acceleration $\dot{\bf v}_{t}$  
(i.e. evaluated at  $t$, as would normally be expected for the dynamics of a particle in classical mechanics), 
but is {\em instead} proportional to the acceleration $[\dot{\bf v}]_{t_{\rm o}}$ at the retarded  time $t_{\rm o}$. 
\begin{equation}
\label{eq:3e}
\frac{{\rm d}{\cal W}}{{\rm d}t}=-\frac{d{\cal E_{\rm me}}}{{\rm d}t}=-{\bf f}_t \cdot {\bf v}_t
= m_{\rm el}\:[\dot{\bf v}]_{t_{\rm o}}\cdot {\bf v}_t=\frac{2e^{2}}{3r_{\rm o} c^{2}}[\dot{\bf v}]_{t_{\rm o}} \cdot {\bf v}_t\:.
\end{equation}
We have shown that if we rewrite force in Eqs.~(\ref{eq:3d1}) and (\ref{eq:3e}) by expressing the retarded value of the acceleration 
in terms of the ``present'' values of the acceleration and its time derivative (Eq.~(\ref{eq:22})), we get the self-force as well as 
the rate of work done against 
the self-force by the charge in terms of the real-time values of quantities specifying the motion of the charge at that 
particular instant (Eqs.~(\ref{eq:3a1}) and (\ref{eq:7}) respectively). 

However if we as such kept the time-retarded value of acceleration in the expression for force 
and instead expressed the velocity also in terms of its 
value at the retarded time ${t_{\rm o}}$ (to the required order in $r_{\rm o}/c$), 
\begin{equation}
\label{eq:21.1}
{\bf v}_t=[{\bf v}_{t_{\rm o}}+\dot{\bf v}_{t_{\rm o}}{r_{\rm o}}/{c}]
\end{equation}
we get, 
\begin{equation}
\label{eq:8}
\frac{d{\cal W}}{dt}=-\frac{{\rm d}{\cal E_{\rm me}}}{{\rm d}t}=m_{\rm el}\:[\dot{\bf v}] \cdot {\bf v}_t
=\frac{2e^{2}}{3r_{\rm o} c^{2}}\:\left[\dot{\bf v}_{t_{\rm o}}
{\bf .v}_{t_{\rm o}} +\dot{\bf v}^{2}_{t_{\rm o}}\frac{r_{\rm o}}{c}\right]\:,
\end{equation}
or
\begin{equation}
\label{eq:9}
\frac{d{\cal W}}{dt}=-\frac{{\rm d}{\cal E_{\rm me}}}{{\rm d}t}= m_{\rm el}[\dot{\bf v}_{t_{\rm o}}\cdot
{\bf v}_{t_{\rm o}}]+\frac{2e^{2}}{3c^{3}}[\dot{\bf v}^{2}_{t_{\rm o}}] \:,
\end{equation}
which encompasses Larmor's formula.
These results become more and more precise as $r_{\rm o}$ is made smaller and smaller. 

Equations (\ref{eq:7}) and  (\ref{eq:9}) apparently look very different. It is a general belief in the literature that Larmor's formula 
yields the instantaneous radiative losses from a charge. But when confronted with a different power loss formula, as derived from the 
radiation reaction, there was a desire to preserve its conformity with Larmor's formula, the latter always thought to be the correct one. 
Schott \cite{7} suggested an ad hoc term that converts the radiative loss formula from Eq.~(\ref{eq:7}) to Eq.~(\ref{eq:9}),
\begin{equation}
\label{eq:9.1}
\frac{2e^{2}}{3c^{3}}\dot{\bf v}^{2}= -\frac{2e^{2}}{3c^{3}}\ddot{\bf v}\cdot{\bf v}
+\frac{2e^{2}}{3c^{3}}\frac{{\rm d}(\dot{\bf v}\cdot{\bf v})}{{\rm d}t}\:.
\end{equation}
The term $2e^{2}\dot{\bf v}\cdot{\bf v}/3c^{3}$ is known in the literature as the acceleration energy or an 
internal energy term or simply a ``Schott-term'', which is added arbitrarily so as to just make the power loss due to 
radiation reaction comply with Larmor's formula for radiation. 
The Schott-term is zero when acceleration is perpendicular to the charge velocity, i.e., when  
$\dot{\bf v}\cdot{\bf v}=0$. Of course in this case $\ddot{\bf v}\cdot{\bf v}+\dot{\bf v}^{2}=0$, and both  
Eqs. (\ref{eq:7}) and  (\ref{eq:9}) yield equal values for the radiative losses.
The meaning of this a-century-old term is still being debated 
and it has not provided a well-accepted solution to the problem.  This internal energy term is elusive as it does not seem to make 
appearance anywhere else. It is not even clear whether a variation in ``internal energy'' term causes a variation in the 
effective mass of the charge. But as we have shown here the two formulas 
(Eqs.~(\ref{eq:7}) and Eq.~(\ref{eq:9})) are compatible without the introduction of any such additional internal energy term and that 
the apparent discordance results from two different time systems being employed in arriving at the two formulas.
The same thing could be realized from the difference in the temporal rate of the change of energy in the self-fields between 
the retarded and real times. From the difference in the first terms on the right hand sides of Eqs. (\ref{eq:7}) and (\ref{eq:9}) we get,
\begin{equation}
\label{eq:9.2}
m_{\rm el}(\dot{\bf v}\cdot {\bf v})_t - m_{\rm el}(\dot{\bf v}\cdot{\bf v})_{t_{\rm o}}
= \frac{2e^{2}}{3r_{\rm o}c^{2}}\frac{{\rm d}(\dot{\bf v}\cdot{\bf v})}{{\rm d}t}\tau_{\rm o}
=\frac{2e^{2}}{3c^{3}}\frac{{\rm d}(\dot{\bf v}\cdot{\bf v})}{{\rm d}t}\:.
\end{equation}
From this it is obvious that the so-called acceleration-dependent internal-energy term is nothing but the difference in the 
energy in self-fields of the charge between the retarded and the present time.
This is also consistent with Schott-term being zero when $\dot{\bf v}\perp{\bf v}$ as then there is no change in the speed of the charge and 
as a result energy in self-fields also does not change between the retarded and the present time.
This should thus obviate the need for this mysterious internal-energy term. 
\section{A Pictorial Demonstration of Relation Between Radiation Reaction and Larmor's Formula}
The genesis of radiative power losses from an accelerated charge can be best understood pictorially for  a charge (a small sphere of vanishingly 
small radius $r_{\rm o}$) undergoing a circular motion. This also demonstrates in a succinct manner the relation between radiation reaction and 
Larmor's formula. Let the charge at some instant $t$ be lying  
at a point A on the circle (Fig. 1), having a velocity ${\bf v}$ and an acceleration $\dot{\bf v}$. The acceleration and velocity are 
related by $\dot{\bf v}=\mbox{\boldmath $\omega$} \times {\bf v}$, where  $\mbox{\boldmath $\omega$}$ is the angular velocity vector. 
The centripetal force on the charge at A is along the radial direction AO, perpendicular to the velocity vector ${\bf v}$ at A, 
therefore there should be no work done by the charge and consequently no change in its kinetic energy. 
\begin{figure}[ht]
\includegraphics[width=\columnwidth]{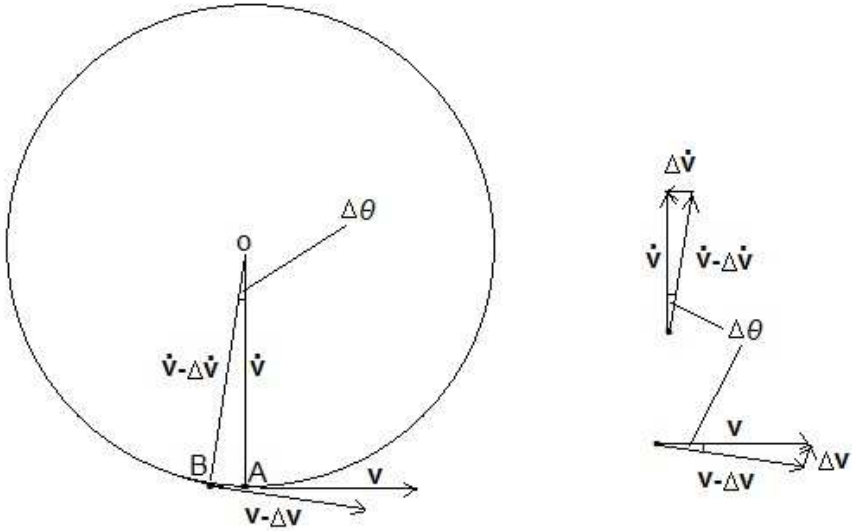}
\caption{Circular motion of a charge moving with velocity $\bf v$ and acceleration $\dot{\bf v}$ at a time $t$ at point A. 
The charge has moved from B to A by an angle $\Delta \theta$ during a small time interval $\tau_{\rm o}=t-t_{\rm o}$, during which the velocity and 
acceleration have changed by $\Delta {\bf v}$ and  $\Delta \dot{\bf v}$ respectively.} 
\end{figure}

Let $\Delta \theta= \omega \tau_{\rm o} $ be the angle, traversed during the small time interval $\tau_{\rm o}$, as the charge moves from say, B to A (Fig. 1).
The distance between points B and A is less than $r_{\rm o}$, though  for clarity they might be shown quite apart in a schematic diagram 
(fig. 1). Then the force on the charge at A is proportional to the acceleration it had at point B which was along BO with  
a component not only along the radial direction AO, but also a tiny component perpendicular to that.
Splitting the acceleration into these two components, $[\dot{\bf v}]_{t_{\rm o}}=\dot{\bf v}-\Delta \dot{\bf v}$, 
the self-force experienced by the charge at A is,
\begin{equation}
\label{eq:3d2}
{\bf f}=-m_{el} [\dot{\bf v}]_{t_{\rm o}}=-m_{el} (\dot{\bf v}-\Delta \dot{\bf v})=
-m_{el}(\dot{\bf v}-\ddot{\bf v}\tau_{\rm o})=-m_{el}\dot{\bf v}+\frac{2e^{2}}{3 c^{3}}\ddot{\bf v}\:,
\end{equation}
which is the formula for radiation reaction (Eq.~(\ref{eq:3a1})) for a non-relativistic motion of the charge. 
The radiative power loss then is 
\begin{equation}
\label{eq:3d3}
P=-{{\rm d}{\cal E}}/{{\rm d}t}= -{\bf f}\cdot {\bf v}=-\frac{2e^{2}}{3c^{3}} (\ddot{\bf v}\cdot {\bf v})\:, 
\end{equation}
since $\dot{\bf v}\cdot {\bf v}=0$ in the circular motion. It is clear that $\Delta \dot{\bf v}=\ddot{\bf v}\tau_{\rm o}$ is in a 
direction opposite to ${\bf v}$ (Fig. 1), 
therefore $-(\ddot{\bf v}\cdot {\bf v})$ is a positive quantity and energy is lost by the charge.

Alternatively one can  express the velocity at A in terms of the time retarded values at B,  
${\bf v}=[{\bf v}+\dot{\bf v}\tau_{\rm o}]_{t_{\rm o}}$, with velocity components perpendicular and parallel to BO, to write the power loss as, 
\begin{equation}
\label{eq:3d4}
P=-{{\rm d}{\cal E}}/{{\rm d}t}=m_{el} [\dot{\bf v} \cdot ({\bf v}+\dot{\bf v}\tau_{\rm o})]_{t_{\rm o}}
=\frac{2e^{2}}{3r_{\rm o} c^{2}} [\dot{\bf v}\cdot \dot{\bf v}]_{t_{\rm o}}\:, 
\end{equation}
which is nothing but Larmor's formula of radiative power loss.

Thus while in the case of radiation reaction the radiative power loss is expressed in terms of the real time motion of the charge, 
the Larmor's formula gives the radiative loss in terms of the charge motion at a retarded time. 
In case of a circular motion, the two formulas yield equal values for the radiative power loss, as expected from our earlier discussion.
There is however a caveat that in a general motion, 
where $\dot{\bf v}\cdot{\bf v} \ne 0$, the two power formulas would not give equal results. 
\section{Maxwell Stress Tensor and the Self-Force}
We examined the consistency of instantaneous rate of work done against the self-forces, with Larmor's formula  
calculated from Poynting flux through surface $\Sigma$, thus verifying the electromagnetic energy conservation 
in the case of an accelerated charge. 
Likewise, we can also verify the electromagnetic momentum conservation for an accelerated charge by examining 
the consistency of the self-force itself (Eq.~(\ref{eq:3a}) 
or equivalently Eq.~(\ref{eq:3d})) on the charge, with the Maxwell stress tensor, the electromagnetic 
momentum flux density. The latter 
lets us calculate the flow of momentum across the boundary surface $\Sigma$ into the enclosed 
volume which in turn represents the force acting on the combined system of particles and fields within the volume \cite{1,25}. 
Thus the force on the charge can be calculated, in an alternative method, by making use of the 
electromagnetic momentum conservation of the system.
From a surface integral of Maxwell stress tensor $T$ on $\Sigma$ at $t={t_{\rm o}}$, 
we write the force on the charge enclosed within surface $\Sigma$ as \cite{1,25},
\begin{equation}
\label{eq:1d}
{\bf f}=\frac{{\rm d}{\cal P}_{\rm me}}{{\rm d}t}= \oint_{\Sigma}{{\rm d}\Sigma}\:{\bf n} \cdot {\bf T}
-\frac{{\rm d}{\cal P_{\rm em}}}{{\rm d}t}\:,
\end{equation}
Here ${\cal P_{\rm em}}=(1/4\pi c)\int{{\rm d}V}\:({\bf E}\times{\bf B})$ is the volume integral of the electromagnetic field momentum 
within the charged sphere, proportional to $r_{\rm o} $ (or its higher powers), and can therefore be neglected 
for a vanishing small $r_{\rm o} $, implying ${\rm d}{\cal P_{\rm em}}/{\rm d}t=0$. 
Then
\begin{equation}
\label{eq:1d1}
{\bf f}= \oint_{\Sigma}\frac{{\rm d}\Sigma}{4\pi}\left[({\bf n} \cdot {\bf E}){\bf E}+({\bf n} \cdot {\bf B}){\bf B}-\frac{1}{2}
{\bf n}(E^2+B^2)\right]\:,
\end{equation}
where ${\bf E}$ and ${\bf B}$, electromagnetic fields on the surface $\Sigma$ of radius $r_{\rm o}$ of the charge 
instantaneously stationary at $t={t_{\rm o}}$, are given by Eqs.~(\ref{eq:2.1}) and (\ref{eq:2b1}).

Now ${\bf n} \cdot {\bf B}=0$  and the last term ($\propto {\bf n}B^2$) on the right hand side, when integrated over 
the surface $\Sigma$ of the whole sphere, yields a zero net value. Substituting for ${\bf E}$ from Eq.~(\ref{eq:2.1}) in Eq.~(\ref{eq:1d1}), 
and dropping terms that either go to zero for a vanishing small $r_{\rm o}$ or yield a zero net value when integrated over 
the surface $\Sigma$ of the whole sphere, we get,
\begin{eqnarray}
\nonumber
{\bf f}= \frac{e^2}{4\pi}\oint_{\Omega} {\rm d}\Omega\left[
\left\{-\frac{2\dot{\bf v}}{3r_{\rm o} c^{2}}+\frac{2\ddot{\bf v}}{3c^{3}}\right\}
+{\bf n}\left\{-\frac{2{\bf n} \cdot \dot{\bf v}}{3r_{\rm o} c^{2}}+\frac{2{\bf n} \cdot \ddot{\bf v}}{3c^{3}}\right\}\right.\;\;\;\;\\
\label{eq:1d2}
-\left.\frac{{\bf n}}{2}\left\{-\frac{4{\bf n} \cdot \dot{\bf v}}{3r_{\rm o} c^{2}}+\frac{4{\bf n} \cdot \ddot{\bf v}}{3c^{3}}\right\}
\right]\:.
\end{eqnarray}
Then from Eq.~(\ref{eq:1d2}) we get,
\begin{equation}
\label{eq:1d3}
{\bf f}= \frac{e^2}{2}\int_{\rm o}^{\pi} {\rm d}\theta\:\sin\theta\left[-\frac{2\dot{\bf v}}{3r_{\rm o} c^{2}}
+\frac{2\ddot{\bf v}}{3c^{3}}\right]\:,
\end{equation}
or
\begin{equation}
\label{eq:1d4}
{\bf f}=-\frac{2e^{2}\dot{\bf v}}{3r_{\rm o} c^{2}}+\frac{2e^{2}\ddot{\bf v}}{3c^{3}}
 =-\frac{2e^{2}}{3r_{\rm o} c^{2}}[\dot{\bf v}_{\rm o}] \:.
\end{equation}

This expression for force on the charge enclosed within the surface $\Sigma$, derived here from the electromechanical momentum 
conservation of the system, is consistent with the self-force of the accelerated charge evaluated from the 
detailed mutual interaction between its constituents.
\section{Radiation Reaction and Radiative Power Losses for a Relativistic Motion of the Charge}
We have seen that the radiation reaction (Eq.~(\ref{eq:3a1})) on an accelerated charged sphere of vanishingly small radius 
$r_{\rm o} $ is nothing but the self-force proportional to the acceleration it had at a time $r_{\rm o}/c$ earlier (Eq.~(\ref{eq:3d1})). 
Till now we considered a charge that was at most moving with a non-relativistic speed. A question may arise -- are our inferences equally
valid for a charge that is moving with a relativistic speed? After all a moving charge would be Lorentz contracted along its direction of 
motion, breaking the spherical symmetry, then what will be the retarded time corresponding to $t_{\rm o}=t-r_{\rm o} /c$ for an instantly 
stationary charge? 

Let ${\cal K}$ be an inertial frame (lab-frame!) in which the charge at some instant $t$ is moving with velocity ${\bf v}$ and acceleration 
$\dot{\bf v}$. Let ${\cal K}'$ be the instantaneously rest frame of the charge at that event. 
The average of the light travel time between 
various infinitesimal constituents of the spherical charge in the rest frame ${\cal K}'$ is $\tau_{\rm o} = r_{\rm o} /c$.
Now in frame ${\cal K}$, due to the time dilation, all time intervals between any pair of charge elements, which are 
instantaneously stationary in frame ${\cal K}'$, will be longer by the Lorentz factor $\gamma=(1-v^2/c^2)^{-1/2}$ and so will be their 
average retarded time. Thus in frame ${\cal K}$ the effective retarded time will be $\tau=\gamma\tau_{\rm o}=\gamma r_{\rm o}/c$. 
This might look surprising considering that the distances along the direction of motion are  
in fact contracted between all charge elements by the Lorentz factor $\gamma$. 

Actually due to the charge motion the average light travel time across the charged sphere would be larger. 
The light travel time $\tau$ from the centre of a moving Lorentz-contracted sphere (ellipsoid) to a point at an angle $\theta$, 
where $\theta$ is the angle with respect to direction of motion, is calculated from,
\begin{equation}
\label{eq:3g1}
\left(\frac {r_{\rm o}\cos \theta}{\gamma}+v\tau\right)^2+r^2_{\rm o}\sin^2 \theta= (c\tau)^2,
\end{equation}
which has a solution $\tau=(1+v \cos \theta/c)\gamma r_{\rm o}/c$. When averaged over all $\theta$, we get for the average light travel 
time $\tau=\gamma r_{\rm o} /c=\gamma\tau_{\rm o}$.
\subsection{Motion of the Charge in One Dimension}
A generalization of Eq.~(\ref{eq:3d1}) to a relativistic case is straightforward in a one-dimensional motion, 
where acceleration is ever parallel to the velocity ($\dot{\bf v} \parallel {\bf v}$). We simply replace $\dot{\bf v}$ 
with the proper acceleration $\gamma^3 \dot{v}$ as this precisely is what we get when we make a relativistic transformation of 
Eq.~(\ref{eq:3d1}) from the instantaneous rest frame ${\cal K}'$ of the charge  to the lab-frame ${\cal K}$ for a one-dimensional motion. 
Then the relativistic counterpart of the radiation reaction Eq.~(\ref{eq:3d1}) is simply, 
\begin{equation}
\label{eq:3f6}
{f}_{t}=-m_{\rm el}[\gamma^3 \dot{v}]_{t_{\rm o}}. 
\end{equation}

The self-force on the charge (with all quantities expressed in terms of their values at real time $t$) is,
\begin{equation}
\label{eq:3f4.1}
{f} = -\frac{2e^2}{3r_{\rm o} c^2}\gamma^3 \dot{v}
+\frac{2e^2}{3r_{\rm o} c^2}\frac{{\rm d}(\gamma^3 \dot{v})}{{\rm d}t}\gamma r_{\rm o} /c
\end{equation}
or
\begin{eqnarray}
\label{eq:3f.1}
{f}= - m_{\rm el}\:\gamma^3 \dot{v}
+\frac{2e^2\gamma ^{4}}{3c^3}\left\{\ddot{v}+\frac{3\gamma ^{2}(\dot{ v}{ v})\dot{v}}{c^{2}}\right\} \:.
\end{eqnarray}
The first term on the right hand side represents the present rate of change of the mechanical momentum of the charge while the remainder 
(i.e., terms enclosed within the curly brackets) represents the radiation reaction. 
The relativistic counterpart of Eq.~(\ref{eq:3e}) is easily obtained from  Eq.~(\ref{eq:3f.1}) by a scalar product with the 
instantaneous velocity of the charge,
\begin{equation}
\label{eq:10c}
\frac{{\rm d}{\cal W}}{{\rm d}t}=-\frac{{\rm d}{\cal E_{\rm me}}}{{\rm d}t}=m_{\rm el}(\gamma^3 \dot{v}{ v})
-\frac{2e^{2}\gamma ^{4}}{3c^{3}}\left\{\ddot{ v}{ v}+\frac{3\gamma ^{2}(\dot{ v}{ v})^{2}}{c^{2}}\right\}.
\end{equation}
The first term on the right hand side represents the (present) rate of change of the kinetic energy ($m_{\rm el}\:c^2{\rm d}\gamma/{\rm d}t$) 
of the electrical mass of the charge 
while the remainder give the relativistic formula for power loss due to radiation reaction.

Now we want to convert the power loss in terms of all quantities expressed at their retarded time values. 
A subtle point to be kept in mind is that while ${\rm d}{\cal E_{\rm me}}/{\rm d}t$ in Eq.~(\ref{eq:7}) as well as (\ref{eq:10c}) provides
rate of kinetic energy change in terms of all quantities for real time, ${\rm d}{\cal E_{\rm me}}/{\rm d}t$ in Eq.~(\ref{eq:9}) gives rate 
of kinetic energy change in terms of the retarded time values, including the infinitesimal time intervals.
In order to relate the two infinitesimal time intervals one can take a cue from the relation between the proper time interval 
${\rm d}t'$ in rest frame ${\cal K}'$ and the corresponding time interval ${\rm d}t=\gamma{\rm d}t'$ in the lab-frame ${\cal K}$ at the 
real times $t'$ and $t$ and a similar relation between time intervals ${\rm d}t'_{\rm o}$ and 
${\rm d}t_{\rm o}=\gamma_{\rm o}{\rm d}t'_{\rm o}$ 
at retarded times $t'_{\rm o}$ and $t_{\rm o}$ in the two respective frames. The time intervals ${\rm d}t'$ and ${\rm d}t'_{\rm o}$ 
are equal in rest frame ${\cal K}$ (that is why we did not distinguish between the time intervals 
in Eqs.~(\ref{eq:7}) and Eq.~(\ref{eq:9}). Then we get,
\begin{equation}
\label{eq:3e1}
{\rm d}t/{\rm d}t_{\rm o}=\gamma/\gamma_{\rm o}=1+(\gamma^2 \dot{\bf v}\cdot{\bf v}/c^2)({\gamma r_{\rm o}}/c)\;,
\end{equation}
Alternately we can differentiate $t=t_{\rm o}+\gamma r_{\rm o} /c$ to get Eq.~(\ref{eq:3e1}).

The velocity ${\bf v}$ can be written in terms of retarded time velocity and acceleration as, 
\begin{equation}
\label{eq:21.2}
{\bf v}_{t}=[{\bf v}_{t_{\rm o}}+\dot{\bf v}_{t_{\rm o}}({\gamma r_{\rm o}}/c)]
\end{equation}
Now from Eqs.~(\ref{eq:3f6}), (\ref{eq:3e1}) and (\ref{eq:21.2}) we calculate power loss rate for the one-dimensional motion as,
\begin{equation}
\label{eq:10d}
\frac{{\rm d}{\cal W}}{{\rm d}t_{\rm o}}=-\frac{{\rm d}{\cal E_{\rm me}}}{{\rm d}t_{\rm o}}= (-f v)_{t}\:\frac{{\rm d}t}{{\rm d}t_{\rm o}}
=m_{\rm el}\left[\gamma^3 \dot{v}{ v}\right]
+\frac{2e^2\gamma ^{4}}{3c^3}\left[\dot{ v}\dot{ v}+\frac{\gamma ^{2}(\dot{ v}{ v})^{2}}{c^{2}}\right].
\end{equation}
The first term on the right hand side represents the rate of change of the kinetic energy ($m_{\rm el}\:c^2{\rm d}\gamma/{\rm d}t$) 
of the electrical inertial mass of the charge 
at the retarded time while the remainder give the relativistic generalization of Larmor's formula for radiative power loss.
\subsection{The Charge with an Arbitrary Motion}
A more general extension to 
${\bf f}_{t}= -m_{\rm el}[\gamma^3 \dot{\bf v}_\parallel + \gamma \dot{\bf v}_\perp]_{t_{\rm o}}$, 
where acceleration has components both parallel and perpendicular to the velocity, does not beget a relativistically covariant formulation.
This is because of certain nuances of relativistic transformation that we need to be wary of in our formulation. 

For instance, while the parallel component of acceleration transforms from the rest frame ${\cal K}'$ to the lab-frame ${\cal K}$ as,
\begin{equation}
\label{eq:3f2.1}
\dot{\bf v}'_\parallel = \gamma^3 \dot{\bf v}_\parallel \:,
\end{equation}
the perpendicular component of acceleration transforms as \cite{71}, 
\begin{equation}
\label{eq:3f2.2}
\dot{\bf v}'_\perp = \gamma^2 \dot{\bf v}_\perp+\gamma^4 (\dot{v}_\parallel {v}/c^2){\bf v}_\perp\:.
\end{equation}
On the other hand the parallel component of force transforms as \cite{71} 
\begin{equation}
\label{eq:3f2.3}
{\bf f}'_\parallel={\bf f}_\parallel - \gamma^2 ({\bf f}_\perp \cdot {\bf v}_\perp) {\bf v}_\parallel /c^2\:,
\end{equation}
while for the perpendicular component of force the transformation is,
\begin{equation}
\label{eq:3f2.4}
{\bf f}'_\perp = \gamma{\bf f}_\perp \:.
\end{equation}
The second terms on the right hand sides of  Eqs.~(\ref{eq:3f2.2}) and (\ref{eq:3f2.3}) make contributions at the retarded time 
$t_{\rm o}=t-\gamma r_{\rm o} /c$ due to the perpendicular velocity component $[{\bf v}_\perp =-\dot{\bf v}_\perp \gamma r_{\rm o}/c]$ 
the charge has because of the acceleration $\dot{\bf v}_\perp$, even though the perpendicular velocity component is zero at $t$ 
in our chosen frame ${\cal K}$. 

Thus for the parallel component of force at real time $t$ we get from Eqs.~(\ref{eq:3f2.1}) and (\ref{eq:3f2.3}),  
\begin{equation}
\label{eq:3f1}
{\bf f}'_\parallel= -m_{\rm el}[\gamma^3 \dot{\bf v}_\parallel]_{t_{\rm o}}\:,
\end{equation}
with the second term on the right hand side of Eqs.~(\ref{eq:3f2.2}) making a nil contribution at $t$, as we discussed above.

For the perpendicular component of force at $t$, Eqs.~(\ref{eq:3f2.2}) and (\ref{eq:3f2.4})  
might appear to lead to ${\bf f}_\perp \propto -\gamma \dot{\bf v}_\perp$, but here for calculating the 
self-force of the charge we need to specify the acceleration at time $t_{\rm o}$.
\begin{eqnarray}
\nonumber
\gamma{\bf f}_\perp = -m_{\rm el}[\gamma^2 \dot{\bf v}_\perp]_{t_{\rm o}}
+ \frac{2e^2}{3r_{\rm o} c^2}\gamma^4 (\dot{v}_\parallel {v}/c^2)(\dot{\bf v}_\perp \gamma r_{\rm o}/c) \\
\label{eq:3f3}
= -m_{\rm el}[\gamma^2 \dot{\bf v}_\perp]_{t_{\rm o}} + \frac{2e^2}{3 c^5}\gamma^5 \dot{v}_\parallel {v}\dot{\bf v}_\perp\:.
\end{eqnarray}

Then components of the self-force on the charge (with all quantities expressed in terms of their values at real time $t$) are,
\begin{equation}
\label{eq:3f4}
{\bf f}_\parallel = -\frac{2e^2}{3r_{\rm o} c^2}\gamma^3 \dot{\bf v}_\parallel
+\frac{2e^2}{3r_{\rm o} c^2}\frac{{\rm d}(\gamma^3 \dot{\bf v}_\parallel)}{{\rm d}t}\gamma r_{\rm o} /c \;,
\end{equation}
and 
\begin{eqnarray}
\label{eq:3f5}
\gamma{\bf f}_\perp = - m_{\rm el}\gamma^2 \dot{\bf v}_\perp
+ \frac{2e^2}{3r_{\rm o} c^2} \frac{{\rm d}(\gamma^2 \dot{\bf v}_\perp)}{{\rm d}t}\gamma r_{\rm o} /c 
+ \frac{2e^2}{3 c^5}\gamma^5 \dot{v}_\parallel {v}\dot{\bf v}_\perp\:.
\end{eqnarray}
Combining the two force components we get,
\begin{eqnarray}
\nonumber
{\bf f}= - m_{\rm el}(\gamma^3 \dot{\bf v}_\parallel+\gamma\dot{\bf v}_\perp)
+\frac{2e^2}{3c^3}\left\{\frac{{\rm d}(\gamma^3 \dot{\bf v}_\parallel)}{{\rm d}t}\gamma
+\frac{{\rm d}(\gamma^2 \dot{\bf v}_\perp)}{{\rm d}t}+\frac{\gamma^4 \dot{v}_\parallel v\dot{\bf v}_\perp}{c^2}\right\}\\
\nonumber
= - m_{\rm el}(\gamma^3 \dot{\bf v}_\parallel+\gamma\dot{\bf v}_\perp)
+\frac{2e^2}{3c^3}\left\{\frac{3\gamma ^{6}(\dot{\bf v}\cdot{\bf v})\dot{\bf v}_\parallel}{c^{2}}
+\gamma ^{4}\ddot{\bf v}_\parallel+\frac{2\gamma^{4}(\dot{\bf v}\cdot{\bf v})\dot{\bf v}_\perp}{c^{2}}\right.\\
\label{eq:3f}
+\left.\gamma^2 \ddot{\bf v}_\perp+\frac{\gamma ^{4}(\dot{\bf v}\cdot{\bf v})\dot{\bf v}_\perp}{c^2}\right\} \:.
\end{eqnarray}
Rearranging various terms we can write,
\begin{eqnarray}
\nonumber
{\bf f}= - m_{\rm el}(\gamma^3 \dot{\bf v}_\parallel+\gamma\dot{\bf v}_\perp)
+\frac{2e^{2}}{3 c^{3}}\left\{\frac{3\gamma ^{6}(\dot{\bf v}\cdot{\bf v})^2{\bf v}}{c^{4}}
+\frac{\gamma ^{4}(\ddot{\bf v}\cdot{\bf v}){\bf v}}{c^{2}}\right.\\
\label{eq:3i}
\left.+\frac{3\gamma ^{4}(\dot{\bf v}\cdot{\bf v})\dot{\bf v}}{c^{2}}+\gamma^2 \ddot{\bf v}\right\}\:.
\end{eqnarray}
The first term on the right hand side represents the present rate of change of momentum of the electrical inertial mass of the charge 
while the remainder (i.e., terms enclosed within the curly brackets) represents the radiation reaction, obtained 
here in a quite simple way but which was obtained in the literature after rather lengthy calculations \cite{7,20} . 
The relativistic counterpart of Eq.~(\ref{eq:3e}) is easily obtained from  Eq.~(\ref{eq:3i}) by a scalar product with the 
instantaneous velocity of the charge,
\begin{equation}
\label{eq:10a}
\frac{{\rm d}{\cal W}}{{\rm d}t}=m_{\rm el}(\gamma^3 \dot{\bf v}\cdot{\bf v})
-\frac{2e^{2}\gamma ^{4}}{3c^{3}}\left\{\ddot{\bf v}\cdot{\bf v}+\frac{3\gamma ^{2}(\dot{\bf v}\cdot{\bf v})^{2}}{c^{2}}\right\}.
\end{equation}
The first term on the right hand side represents the (present) rate of change of the kinetic energy 
($m_{\rm el}\:c^2{\rm d}\gamma/{\rm d}t$) 
of the electrical inertial mass of the charge while the remainder gives the relativistic formula for power loss due to radiation reaction.

Thus the general formula for radiative losses from a charge is,
\begin{equation}
\label{eq:10}
P=-\frac{2e^{2}\gamma ^{4}}{3c^{3}}\left\{\ddot{\bf v}\cdot{\bf v}+
\frac{3\gamma ^{2}(\dot{\bf v}\cdot{\bf v})^{2}}{c^{2}}\right\}\:.
\end{equation}

Now the parallel component of force at the retarded time $t_{\rm o}$ is written from Eqs.~(\ref{eq:3f2.3}) as,
\begin{equation}
\label{eq:3f7.1}
[{\bf f}_\parallel]= -m_{\rm el}[\gamma^3 \dot{\bf v}_\parallel]_{t_{\rm o}}
+\gamma^2 ({\bf f}_\perp \cdot \dot{\bf v}_\perp) {\bf v}_\parallel \gamma r_{\rm o} /c^3 \:,
\end{equation}
while for the perpendicular component we get,
\begin{equation}
\label{eq:3f7.2}
[{\bf f}_\perp]= -m_{\rm el}[\gamma \dot{\bf v}_\perp]_{t_{\rm o}}
+ \frac{2e^2}{3 c^5}\gamma^4 \dot{v}_\parallel {v}\dot{\bf v}_\perp\:.
\end{equation}
It should be noted that in our case where the forces being considered are self-forces, the sign of the 
second term on the right hand side in Eqs.~(\ref{eq:3f7.1}) reverses from that of Eqs.~(\ref{eq:3f2.3}) as the latter 
has been derived for the normal dynamics of a particle in classical mechanics where the force in consideration is directly 
proportional to the applied acceleration without any negative sign in front.

From Eqs.~(\ref{eq:3f7.1}) and Eq.~(\ref{eq:3f7.2}) the net force can then be written as,
\begin{equation}
\label{eq:3f7}
[{\bf f}]= -m_{\rm el}[\gamma^3 \dot{\bf v}_\parallel+\gamma \dot{\bf v}_\perp]_{t_{\rm o}}
+\frac{2e^2}{3 c^5}\gamma^4 [-(\dot{\bf v}_\perp \cdot \dot{\bf v}_\perp) {\bf v}_\parallel
+ \dot{v}_\parallel {v}\dot{\bf v}_\perp]\:,
\end{equation}
Now from Eqs.~(\ref{eq:3e1}), (\ref{eq:21.2}) and (\ref{eq:3f7}) we calculate power loss rate as,
\begin{eqnarray}
\nonumber
\frac{{\rm d}{\cal W}}{{\rm d}t_{\rm o}}= [-{\bf f}\cdot {\bf v}_{t}] {\rm d}t/{\rm d}t_{\rm o} 
=m_{\rm el}\left[\gamma^3 \dot{\bf v}\cdot{\bf v}\right]
+\frac{2e^2}{3c^3}\left[\gamma ^{4} \dot{\bf v}_\parallel \cdot\dot{\bf v}_\parallel
+\frac{\gamma ^{6}(\dot{\bf v}\cdot{\bf v})^{2}}{c^{2}}\right.\\
\label{eq:10b.1}
\left.+\gamma ^{2}\dot{\bf v}_\perp \cdot\dot{\bf v}_\perp
+\frac{\gamma ^{4}(\dot{\bf v}_\perp \cdot\dot{\bf v}_\perp)({\bf v}\cdot{\bf v})}{c^{2}}\right].
\end{eqnarray}
Here we have dropped terms which become negligible as $r_{\rm 0}\rightarrow 0$. Rearranging various terms, we can write it as,
\begin{equation}
\label{eq:10b}
\frac{{\rm d}{\cal W}}{{\rm d}t_{\rm o}}=m_{\rm el}\:c^2\left[\frac{{\rm d}\gamma}{{\rm d}t_{\rm o}}\right]
+\frac{2e^2\gamma ^{4}}{3c^3}\left[\dot{\bf v}\cdot\dot{\bf v}+\frac{\gamma ^{2}(\dot{\bf v}\cdot{\bf v})^{2}}{c^{2}}\right].
\end{equation}
The first term on the right hand side represents the rate of change of the kinetic energy 
of the electrical inertial mass of the charge at the retarded time while the remainder gives the relativistic generalization of 
Larmor's formula for radiative power loss, which of course is Li\'{e}nard's formula \cite{1,25}, 
\begin{equation}
\label{eq:11}
P=\frac{2e^{2}\gamma ^{4}}{3c^{3}}\left[\dot{\bf v}^{2}+
\frac{\gamma ^{2}({\bf v}\cdot\dot{\bf v})^{2}}{c^{2}}\right].
\end{equation}

From the difference of the first terms on the right hand sides of Eqs. (\ref{eq:10a}) and (\ref{eq:10b})
We can find the difference in the temporal rate of energy going into self-fields of the charge between the retarded and real time. 
\begin{equation}
\label{eq:12}
\Delta P=m_{\rm el}\:c^2\frac{{\rm d}\gamma}{{\rm d}t}
-m_{\rm el}\:c^2\left[\frac{{\rm d}\gamma}{{\rm d}t_{\rm o}}\right]
\end{equation}
However we must convert the second term to the real time units by multiplying it with  
${\rm d}t_{\rm o}/{\rm d}t=\gamma_{\rm o}/\gamma$ from Eq.~(\ref{eq:3e1}).
Then we get,
\begin{equation}
\label{eq:12}
\Delta P=\frac{2e^{2}}{3r_{\rm o}c^{2}}\frac{{\rm d}(\gamma^4\dot{\bf v}\cdot{\bf v})}{\gamma{\rm d}t}\gamma r_{\rm o}/c
=\frac{2e^2\gamma ^{4}}{3c^3}\left[\dot{\bf v}^2+\ddot{\bf v}\cdot{\bf v}
+\frac{4\gamma ^{2}(\dot{\bf v}\cdot{\bf v})^{2}}{c^{2}}\right]
\end{equation}
which indeed is the difference of power loss rate in Eqs. (\ref{eq:11}) and (\ref{eq:10})) that gave rise to   
the ad hoc introduction of relativistic Schott-term ($2e^{2}\gamma^4\dot{\bf v}\cdot{\bf v}/3c^{3}$). 
This confirms the redundancy of this acceleration--dependent internal energy term even in a relativistic case. 
\section{Discussion}
If we examine the rate at which energy is pouring into the 
electromagnetic fields of a charged particle at some given instant, 
then Eq.~(\ref{eq:7}) (Eq.~(\ref{eq:10a}) in a relativistic case)
gives the rate in terms of the real-time values of 
quantities specifying the motion of the charge at that particular 
instant. On the other hand, Eq.~(\ref{eq:9}) 
yields the familiar Larmor's radiation formula 
(Li\'{e}nard's formula from Eq.~(\ref{eq:10b}) in a relativistic case)
but at a cost that a real-time rate of energy going into self-fields of the 
charge distribution is expressed in terms of quantities 
specifying the motion of the charge at a retarded time, 
and not in terms of the quantities specifying the motion at the present time. 
The time difference $\tau_{\rm o}$ may be exceedingly small (infinitesimal 
in the limit $r_{\rm o}\rightarrow 0$), but still its effect on the energy 
calculations is finite because of the presence of the $1/r_{\rm o}$ term for 
energy in the self-fields external to the sphere of radius $r_{\rm o}$. 
Now $\tau_{\rm o}$ is actually the time taken
for a signal to reach from the centre of the sphere 
to the points at its surface. Essentially it implies that if the
electromagnetic energy outflow through the boundary of the spherical
charge distribution (as inferred from the rate of 
work done against the self-force 
of the charge distribution) is expressed in terms of the 
quantities describing the {\em retarded motion} of an
equivalent ``point'' charge at the centre of the sphere,
we obtain the familiar Larmor's formula. We may also point out
that even in the standard text-book statement of
Larmor's formula for radiation from a non-relativistically moving
point charge (see e.g.,~ \cite{1}), the rate of energy flow at a 
time $t$ through a spherical surface of radius $r_{\rm o}$ 
is always expressed in terms of the retarded value of the acceleration
(at time $t_{\rm o}=t-r_{\rm o}/c$) of the point charge, the expression being 
exactly equal to the second term on the right hand side of Eq.~(\ref{eq:9}).  

Equation~(\ref{eq:8}) makes it amply clear why in the instantaneous rest-frame with ${\bf v}=0$, there 
is no radiative loss from the charge. The time-retarded value of velocity will then be 
${\bf v}_{t_{\rm o}}=-\dot{\bf v}_{t_{\rm o}}r_{\rm o}/c$, to a first order, with the first term on the right hand side 
in Eq.~(\ref{eq:8}) yielding $-2e^{2}\dot{\bf v}^2_{t_{\rm o}}/{3 c^{3}}$ and thereby exactly 
cancelling the second term, which otherwise would represent Larmor's result of a finite radiation from an accelerated 
charge even when it is instantly stationary. Also, in the case of a {\em uniformly accelerated} charge, Eq.~(\ref{eq:7}) 
conforms with the conclusions otherwise arrived at that except for the power going into the 
self-fields of the accelerating charge there is no radiation \cite{17,18}. On the other hand Larmor's result of finite radiated power 
has to be taken in conjunction with the first term in Eq.~(\ref{eq:9}), which 
gives power going into the self-fields of the moving charge, but at a rate which is 
not the instantaneous value but only a time-retarded value. It shows that 
the two formulations are mathematically compatible, and either of them could be derived from the other one. 

However a compatibility between them does not necessarily imply that the two formulas are identical and one may need to be  careful 
in their usage. In particular, as we just saw, Larmor's formula could lead to potentially wrong conclusions 
for radiative loss from a uniformly accelerated charge or in the instantaneous rest-frame of a charge with a non-uniform motion.
There is an even more practical case of synchrotron radiation where the two formulas lead to a finite difference in the charge dynamics. 
Using Larmor's formula (or rather its relativistic generalization Li\'{e}nard's formula, viz. (Eq.~(\ref{eq:11})) \cite{1,2,3}) 
it has been concluded that for a relativistic charge spiraling in a uniform magnetic field the pitch angle 
(the angle that the velocity vector makes with the magnetic field direction) does not change in spite of the radiation losses \cite{23}. 
The formulation of the energy losses and radiative life times of gyrating electrons derived thereby has become a standard textbook 
material \cite{26,27}. However it has been shown recently that these conclusions about the constancy of pitch angles are 
inconsistent with the special theory of relativity and that the correct formulation of the pitch angle changes is obtained when radiative  
losses for the gyrating charges are calculated from the radiation-reaction formula \cite{31c}.  

This not only resolves the century--old apparent 
``discrepancy'' in the two power formulas, but also shows an intimate 
relation between the energy in the radiation fields and that in the 
Coulomb self-fields. In particular, the factor of 
4/3 in the electromagnetic mass ($m_{\rm el}=4U_{\rm el}/3c^2$)
of a spherical charge in classical electrodynamics \cite{29}
is intimately connected with the factor of 2/3 found both in 
Larmor's formula and in the radiation-reaction formula. For long, 
this ``mysterious'' factor of 4/3 has been considered to be
undesirable and modifications in classical electromagnetic theory
have been suggested to get rid of this factor (see e.g., \cite{1} and the references
there in). If one does adopt such a modification, then 
one cannot fathom the relation between Larmor's formula and the 
radiation-reaction equation. Moreover, as has been explicitly shown in \cite{15}, the factor of 4/3 in
the electromagnetic inertial mass arises naturally in the conventional electromagnetic theory 
when a full account is taken of all the energy-momentum contributions of   
the electromagnetic forces during the process of attaining the motion of the charged sphere.
Basically there occurs contribution to energy-momentum 
of the charge due to the unbalanced forces of Coulomb self-repulsion of the charge \cite{15}. 
A similar inclusion of the energy-momentum arising from 
the binding forces in the case of a charged capacitor (a macroscopic system!) was shown to lead to a correct explanation~\cite{58}, 
based on energy-momentum arguments, of the null results of the famous Trouton-Noble experiment \cite{59}. 
Of course in an actual charged particle, 
there must be some non-electrical (!) ``binding'' forces (Poincar\'{e} stresses~ \cite{34}) to balance the Coulomb self-repulsion 
of the charge, which would remove the factor of 4/3 in a more natural manner. But these {\em non-electrical} binding forces would not  
enter the expressions for the {\em electromagnetic} fields, thereby leaving the electromagnetic self-fields or the radiation formulas 
intact.  Therefore any such extraneous attempt to modify the classical electromagnetic theory itself, in order to eliminate this factor 
of 4/3, would logically be a step in a wrong direction. 
\section{Conclusions}
We have shown that an accelerated charged particle, comprising a spherical shell of vanishingly small radius, experiences 
at any time a self-force proportional to the acceleration it had earlier, by an instant that a 
light signal takes from the centre of the charged sphere to its surface. 
Accordingly, the instantaneous rate of work done in order to accelerate the charge, obtained by a scalar product of the self-force with 
the ``present'' velocity of the charge, 
is proportional not to the real-time value of the acceleration,  
but instead to a time-retarded value of the acceleration of the charge. 
This rate of work done, when written   
in terms of the real-time value of the acceleration, comprises an additional term proportional to the  
time derivative of the acceleration. Alternatively, the rate of work done at any instant can as well be expressed in terms of 
the square of acceleration, as in Larmor's formula for power radiated,  however, at a cost that the formulation encompasses  
the time-retarded values of the velocity and acceleration of the charge.
In this way, we showed that the rate of work done against the self-forces of a charge 
and Larmor's formula for radiative losses are 
consistent with each other, the apparent discrepancy showing up only due to the fact that while the former 
is written in terms of the instantaneous values of quantities specifying the charge motion, the latter is expressed normally in 
terms of the time-retarded values. There is no need for the acceleration-dependent internal-energy term that was introduced in the 
literature on an ad hoc basis with a desire to make the power loss due to radiation reaction comply with Larmor's formula for radiation.
{}
\end{document}